\documentclass[aps,prl,reprint,superscriptaddress]{revtex4-2}
\usepackage{amsmath,amssymb,graphicx}
\usepackage[colorlinks=true,allcolors=blue]{hyperref}
\begin{document}
\title{Kelvin-Wave Multiplication by Sound Emission}
\author{Weican Yang}
\email{weicanyang@outlook.com}
\affiliation{Yukawa Institute for Theoretical Physics, Kyoto University, Kyoto 606-8502, Japan}
\begin{abstract}
Kelvin waves are bending excitations of a quantized vortex. We identify a subsonic decay channel in which one Kelvin quantum emits a phonon and splits into two lower-energy Kelvin quanta, reducing the Kelvin-wave energy while increasing the number of quanta. For sound wavelengths much longer than the Kelvin wavelengths, rotational symmetry fixes the leading shear contribution to the emission amplitude in terms of the released Kelvin energy. This contribution is odd in the phonon momentum along the vortex. Microscopic calculations support the relation. Coherently seeded Gross--Pitaevskii evolution in a reflecting cavity demonstrates phase-controlled Kelvin gain, with increased mean-square vortex displacement and reduced mean-square slope.
\end{abstract}

\maketitle

Kelvin waves connect vortex motion to sound emission in low-temperature superfluids~\cite{vinen,kozik2004,kozik,krstulovic,proment}. Co-propagating Kelvin excitations pose a distinct relaxation problem: in the long-wavelength subsonic regime, their energy is insufficient for sound to carry away all of their axial momentum. Oppositely propagating quanta can annihilate in pairs~\cite{kozik,horn}; after those partners are depleted, the remaining waves must retain axial momentum while cooling. Kelvin modes and damping have been resolved in trapped condensates~\cite{bretin,mizushima}, and Gross--Pitaevskii (GP) simulations have found radiative transfer towards longer Kelvin wavelengths~\cite{simulaprl,simula}. Ochi et al. calculated axisymmetric phonon emission, $K\to K+B_0$, which preserves Kelvin number and has a soft-emission threshold where the Kelvin group velocity reaches the cylinder's axial sound speed~\cite{ochi}.

Here we identify a subsonic channel in which one Kelvin quantum emits a phonon and splits into two lower-energy Kelvin quanta. The daughters share the retained momentum, while the sound's positive angular momentum permits Kelvin number to increase. For sound wavelengths much longer than the Kelvin wavelengths, rotational invariance relates the leading shear contribution to the emission amplitude, odd in axial phonon momentum, to the parent--daughter energy difference. The linear spectrum thus predicts the daughter-momentum dependence of this part of the four-wave coupling, independently of a direct scattering calculation. It supplies a consistency condition for matching vortex-line effective theories to microscopic models. We test this relation microscopically and demonstrate appreciable coherent conversion in a reflecting cavity. The initial sound phase selects Kelvin amplification or attenuation, accompanied by changes of mean-square vortex displacement and slope.

Consider a straight vortex of positive circulation, described by the condensate field $\psi$ in the GP model,
\begin{equation}
 i\partial_t\psi=(-\nabla^2+|\psi|^2-1)\psi.
 \label{eq:gp21}
\end{equation}
Lengths are in healing lengths $\xi=\hbar/\sqrt{2m_a\mu}$, energies in $\mu=gn_0$, and times in $\hbar/\mu$; $m_a$, $n_0$, and $g$ denote atomic mass, bulk density, and contact coupling. The background is $f(r)e^{i\varphi}$, with $f(\infty)=1$, in cylindrical coordinates $(r,\varphi,z)$. A fundamental Kelvin mode $K_k$ has positive energy $E_K(k)=E_K(-k)$ and angular index $m=-1$ relative to this background~\cite{fetter1965,ochi,papoutsis}. For axial period $L$, define the mode-normalization factor $F=2\pi L N_\xi$, with $N_\xi=n_0\xi^3$. For spontaneous splitting the sound field is initially empty; for stimulated conversion we prepare coherently occupied Kelvin and sound modes. The Kelvin number is $\mathcal N_K=\sum_k n_k$, where $n_k$ is the absolute quantum occupation above the quasiparticle vacuum.

\begin{figure}[t]
\includegraphics[width=\columnwidth]{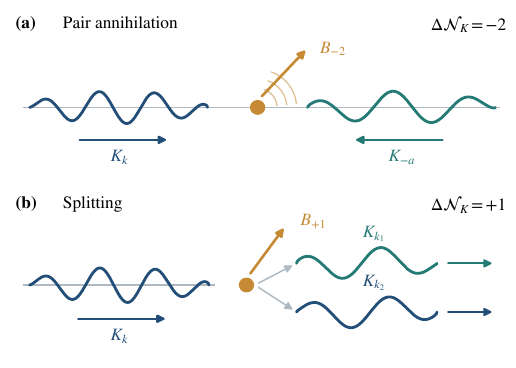}
\caption{Two radiative processes on the same vortex. Counterpropagating Kelvin quanta annihilate into $m=-2$ sound. Splitting leaves two softer quanta and emits $m=+1$ sound, whose positive angular momentum permits an increase of Kelvin number. All curves represent displacement modes of one vortex.}
\label{fig:processes21}
\end{figure}

The two channels in Fig.~\ref{fig:processes21} change the Kelvin number in opposite directions:
\begin{equation}
 \begin{aligned}
 K_k+K_{-a}&\longrightarrow B_{-2},&\Delta\mathcal N_K&=-2,\\
 K_k&\longrightarrow K_{k_1}+K_{k_2}+B_{+1},&\Delta\mathcal N_K&=+1.
 \end{aligned}
 \label{eq:channels21}
\end{equation}
Here $k,a>0$, and $B_m$ denotes sound of angular index $m$. Splitting releases energy $\Omega=E_K(k)-E_K(k_1)-E_K(k_2)>0$ and axial momentum $q=k-k_1-k_2$ to sound. Here subsonic means $E_K'(u)<c=\sqrt2$ throughout $0<u\le k$. Since $E_K(0)=0$, a phonon carrying all the parent momentum costs at least $ck>E_K(k)$. Leaving one Kelvin quantum at $0<k'<k$ also fails: $E_K(k)-E_K(k')=\int_{k'}^k E_K'(u)du<c(k-k')$, below the sound cost. To see how two daughters resolve this restriction, approximate the slowly varying Kelvin logarithm by a constant, $E_K(k)\simeq\Lambda k^2$. For $k_1+k_2\simeq k$, their energy is $\Lambda k^2-2\Lambda k_1k_2$; equal daughters need only half the parent energy. They retain nearly all its momentum and release the difference as sound. The GP dispersion is convex in the interval considered, and the allowed daughters co-propagate with the parent~\cite{sm}.

Angular-momentum balance explains why this channel adds Kelvin-wave action, $\hbar\mathcal N_K$. Each Kelvin quantum carries $-\hbar$ of excess angular momentum: one parent has $-\hbar$, while two daughters have $-2\hbar$, requiring $+\hbar$ in sound. This angular momentum is independent of the axial propagation direction. With the background and angular momentum in other vortex-bound modes fixed,
\begin{equation}
 \mathcal N_K-\mathcal N_{K,0}=\mathcal J_B,
 \label{eq:angular21}
\end{equation}
where $\mathcal J_B$ is the net angular momentum transferred to sound in units of $\hbar$. 

The conservation laws allow the splitting process; its strength is set by the Kelvin--sound emission amplitude. To obtain the constraint imposed by rotational invariance, write the vortex position as $w(z)=x(z)+iy(z)$. Expressing an $xz$ deformation at fixed laboratory $z$ gives $\delta w=\delta x-w'\delta z$: the second term accounts for the change of axial coordinate. In the $xz$ plane, rotation gives $(\delta x,\delta z)=(\varepsilon z,-\varepsilon x)$, whereas symmetric shear gives $(\varepsilon z,+\varepsilon x)$. Thus $\delta w_{\rm rot}=\varepsilon z+\varepsilon xw'$ and $\delta w_{\rm sh}=\varepsilon z-\varepsilon xw'$. Let $H_2,H_4$ be the quadratic and quartic Kelvin energies. At cubic order in $w$, rotational invariance requires the tilt variation of $H_4$ to cancel the coordinate-correction variation of $H_2$~\cite{horn,sonin}. Shear reverses the coordinate correction, so the two contributions add. A long sound wave carries the vortex almost uniformly; in the local fluid frame, this common motion disappears. The Kelvin modes then couple to the transverse--axial strain produced by the longitudinal phonon~\cite{sm}. The deformation acts on all three modes, combining their quadratic energies into the parent energy minus the daughter energies.

Let $s>0$ be the transverse sound wave number and $p=\sqrt{q^2+s^2}$ its magnitude. We take the sound wavelength longer than both daughter Kelvin wavelengths, $p\ll\min(k_1,k_2)$. Since $|q|\le p$, axial recoil only weakly changes the Kelvin wave numbers. At leading order write them as $\alpha_0 k$ and $(1-\alpha_0)k$, with $0<\alpha_0<1$. The released Kelvin energy is
\begin{equation}
 \Delta E_0=E_K(k)-E_K(\alpha_0k)-E_K[(1-\alpha_0)k].
 \label{eq:defect21}
\end{equation}
The sound energy is $\Omega=p\sqrt{p^2+2}$. We neglect recoil in $\Delta E_0$ but retain the sound direction $q/p$: the first nonzero shear response involves sound gradients. A long-wavelength phonon can carry a finite fraction of the parent Kelvin energy. The leading shear contribution to the emission amplitude is
\begin{equation}
 \mathcal M_{\rm sh}=-\sqrt2\,k\Delta E_0\frac{q}{p}
                              \frac{s^{3/2}}{\sqrt\Omega}.
 \label{eq:shear21}
\end{equation}
For distinct daughter modes the matrix element is $\mathcal M_{\rm sh}/F$. The factor $q/p$ comes from longitudinal polarization, $s$ from the transverse sound gradient, and $\sqrt{s/\Omega}$ from phonon normalization. Equation~\eqref{eq:shear21} fixes the leading component odd under $q\to-q$ in the stated soft limit. Density coupling and higher gradients give subleading corrections to this component; the full finite-wave-number amplitude includes other sound couplings~\cite{sm}. For $E_K(k)=k^2[\ln(1/|k|)+C]$, with core constant $C$, equal daughters give $2\Delta E_0/k^2=E_K(k)/k^2-\ln2$. A Kelvin wave disturbs circulating flow over a distance set by its wavelength. Longer-wavelength daughters sample different flow scales, giving this finite subtraction and a coupling that depends on all three energies.

For $E_K(k)\simeq k^2\Lambda(k)$, with $\Lambda(k)\simeq\ln(1/|k|)$, the spontaneous splitting rate per parent scales as $\Gamma_{13}\propto k^{13}\Lambda^5$; the subscript denotes one incoming and three outgoing excitations. Its leading angular pattern is $|Y_{2,1}|^2$, where $Y_{2,1}$ is the spherical harmonic with $\ell=2,m=1$. For sound emitted along the vortex axis, the transverse gradient vanishes; for emission perpendicular to the axis, the axial strain vanishes. Both directions therefore suppress the mixed-shear amplitude. The thin-vortex derivation applies to a zero-temperature, single-component Galilean superfluid with bulk phonons and no additional low-energy core resonance~\cite{horn}.

\begin{figure*}[t]
\includegraphics[width=\textwidth]{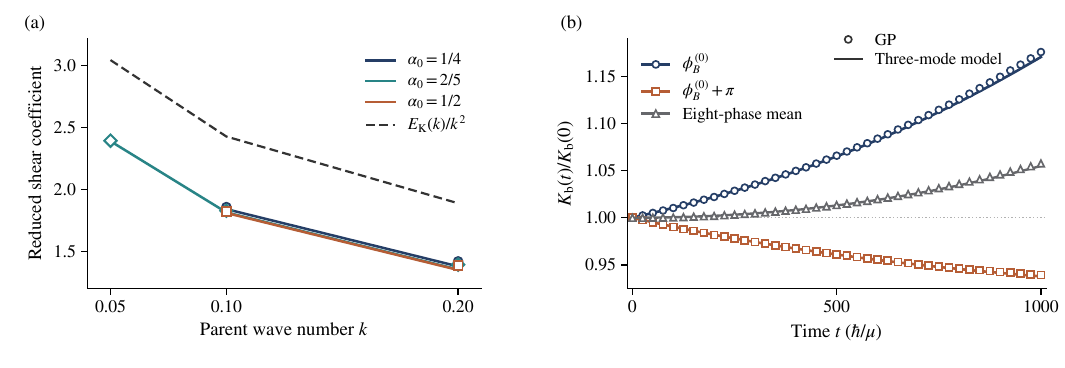}
\caption{Microscopic test of the soft-shear relation and finite-wave-number cavity conversion. (a) Reduced shear coefficient $\Lambda_{\rm odd}^{\rm GP}$ from microscopic GP matrix elements at $q/p=\pm0.5$ (symbols; paired extraction from two energy-conserving sound configurations, defined in Supplemental Material), compared with the dispersion prediction $\Lambda_E$ (solid lines). The dashed estimate $E_K(k)/k^2$ uses only the parent dispersion. (b) Finite-wave-number conversion: total occupation of all retained bound Kelvin modes from GP symbols and independent three-mode predictions for initial sound phases $\phi_B^{(0)}$ and $\phi_B^{(0)}+\pi$. Gray triangles show $\langle K_{\rm b}(t)\rangle_{\phi_B}/\langle K_{\rm b}(0)\rangle_{\phi_B}$ with equal weights for eight initial phases; the gray line is the corresponding phase-averaged three-mode prediction.}

\label{fig:theory21}
\end{figure*}

The GP four-wave scattering amplitude contains the direct quartic interaction and three exchange diagrams with two cubic vertices~\cite{sm,ristivojevic}. Removing the momentum and normalization factors in Eq.~\eqref{eq:shear21} gives the reduced coupling $\Lambda_E=\Delta E_0/[2\alpha_0(1-\alpha_0)k^2]$. Pairing microscopic amplitudes at opposite $q$ extracts the odd shear contribution, which approaches this prediction as $k$ decreases [Fig.~\ref{fig:theory21}(a)]. The parent-only estimate replaces $\Lambda_E$ by $E_K(k)/k^2$, overestimating it by $27$--$40\%$ at the plotted points. Figure~S1 compares the outgoing GP sound with the full microscopic emission amplitude~\cite{sm}. A density-gradient interaction changes the core and dispersion and gives similar agreement with the dispersion-based amplitude~\cite{sm,fu2003,collin2007}.

The spontaneous process is slow; initially occupying the daughter and sound modes makes the same conversion channel appreciable at finite wave number. In a radially reflecting cylinder, periodic along $z$, we prepare a parent at $k_P=0.60$, a daughter at $k_D=0.25$ and $m=+1$ sound at $q=0.10$. Each event creates two quanta in the daughter mode, so $k_P=2k_D+q$. Write the discrete Bogoliubov amplitudes as $a_j=|a_j|e^{i\phi_j}$, $j=P,D,B$, and their occupations divided by $F$ as $I_j=|a_j|^2$; the coherent quantum occupations are $FI_j$. The interaction energy divided by $\mu F$ is $\operatorname{Re}[M_{\rm c}a_P(a_D^*)^2a_B^*]$. Its relative phase is $\phi_P-2\phi_D-\phi_B$. The initial transfer rate is proportional to its sine; a $\pi$ shift of sound phase reverses the rate. Here $M_{\rm c}$ is the full finite-wave-number coupling of the discrete cavity mode, including terms beyond $\mathcal M_{\rm sh}$. The three-mode prediction uses this microscopic coefficient and the occupation-dependent frequency shifts calculated before GP evolution~\cite{sm}. We tune the radius to the initial resonance $\widetilde E_P=2\widetilde E_D+\widetilde E_B$, where tildes include these shifts. The subsequent evolution is unforced; changing occupations can detune the resonance. GP dynamics also includes excitation outside these three modes.

Figure~\ref{fig:theory21}(b) compares preparations with equal initial energy and conserved charges but sound phases differing by $\pi$. Projecting the evolving GP fluctuation onto the stationary vortex's linear normal (Bogoliubov) modes gives $I_\nu=|a_\nu|^2$. We include all retained fundamental Kelvin modes in both axial directions, with wave numbers $q_\nu$, and define $K_{\rm b}=\sum_\nu I_\nu$ and $E_{\rm b}=\sum_\nu E_K(q_\nu)I_\nu$. These are their total occupation divided by $F$ and quadratic energy divided by $\mu F$. By $t=1000$, $K_{\rm b}$ increases by $17.6\%$ and $E_{\rm b}$ decreases by $11.9\%$ for one phase; the opposite phase gives $-6.1\%$ and $+4.3\%$. Total GP energy is conserved: sound, other modes and interaction energy balance the quadratic Kelvin change. Angular-momentum and resolution checks are given in the Supplemental Material~\cite{sm}.

We also compare the complex daughter-amplitude difference between the phase choices. After removing condensate and common linear phases, the three-mode prediction and GP signal differ by $3.9\%$ in the full-trajectory relative $L^2$ norm~\cite{sm}. Coherent vortex--sound driving has been studied previously~\cite{parker,parker2012}; the conversion studied here changes the total retained Kelvin occupation.

Eight GP evolutions with equally spaced initial sound phases and identical initial mode occupations give a $5.8\%$ increase of $K_{\rm b}$ at $t=1000$ [Fig.~\ref{fig:theory21}(b)]. In the three-mode model, phase averaging cancels the initial transfer rate; for these occupations, the positive curvature makes the mean Kelvin occupation initially grow as $t^2$~\cite{sm}.

\begin{figure*}[t]
\includegraphics[width=\textwidth]{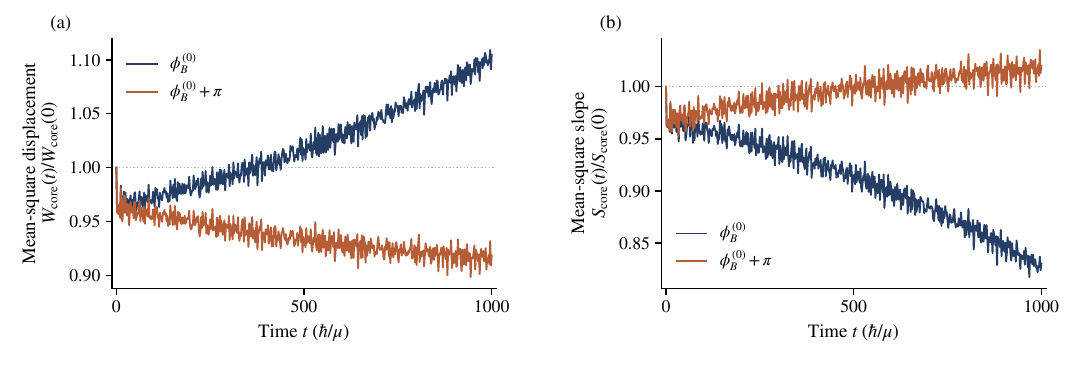}
\caption{Vortex geometry in the same two GP evolutions as Fig.~\ref{fig:theory21}(b). (a) Mean-square displacement of the instantaneous vortex-core position. (b) Mean-square slope. Both quantities are normalized to their own values at $t=0$. The traces use instantaneous core positions and retain the preparation transient, fast oscillations, and acoustic deformation. The initial sound phase reverses the net displacement and slope changes over the conversion interval.}

\label{fig:geometry21}
\end{figure*}

For a helical displacement $w(z)=a_{\rm h}e^{ikz}$, the mean-square displacement is $|a_{\rm h}|^2$, while the mean-square slope is $k^2|a_{\rm h}|^2$. A larger, longer-wavelength excursion can thus vary more gently along the line. Occupation and geometry consequently have different wave-number weights. Let $W_K$ and $S_K$ denote the vacuum-subtracted mean-square displacement and slope of fundamental Kelvin modes after sound has left the core. Then
\begin{equation}
 W_K=\frac1F\sum_kQ(k)n_k,\qquad
 S_K=\frac1F\sum_k k^2Q(k)n_k.
 \label{eq:geometry21}
\end{equation}
The weight $Q(k)$ follows from the Bogoliubov mode at the vortex zero~\cite{sm,martikainen}. A long-wavelength Kelvin mode approaches a rigid vortex translation, giving $Q(k)\to2$, so $W_K\simeq2\mathcal N_K/F$ at long wavelengths. For a given mean-square displacement, longer waves carry less energy and produce a smaller mean-square slope. 

In the cavity, sound and nonlinear deformations also affect vortex position. We determine the line from $\psi[X(z),Y(z),z]=0$, write $w_{\rm core}(z)=X(z)+iY(z)$, and evaluate $W_{\rm core}=L^{-1}\int|w_{\rm core}|^2dz$ and $S_{\rm core}=L^{-1}\int|\partial_z w_{\rm core}|^2dz$. At $t=1000$, the amplifying preparation increases mean-square displacement by $10.5\%$ and reduces mean-square slope by $17.0\%$ [Fig.~\ref{fig:geometry21}]. Shifting the initial sound phase by $\pi$ gives $-8.2\%$ and $+2.0\%$. The corresponding vortex motion is shown in Supplemental Movie S1~\cite{sm}.

For an illustrative $^{87}$Rb condensate with scattering length $100a_0$ and $\mu/h=1\,$kHz, $a_0$ denoting the Bohr radius, $t=1000$ corresponds to $159\,$ms. Let $w_D^\pm(t)$ be the complex Fourier coefficients of vortex displacement at $k_D$ in the two preparations. At $t=1000$, the modulus of their difference is $\xi|w_D^+-w_D^-|=45.7\,$nm~\cite{sm}. The simulated $30.3\,\mu$m axial cell is periodic. A finite experiment would require a longer uniform segment, coherent preparation and control of end reflections over the conversion interval.

For an initially empty open sound continuum, the spontaneous decay rate of one parent at $k=0.60$, $L=50\pi$ and $N_\xi=20$ is $2.93\times10^{-11}\mu/\hbar$~\cite{sm}. Conservative scattering can regenerate reverse-moving partners and restore pair annihilation~\cite{laurie}; reabsorption of reflected sound and thermal drag also enter confined-fluid relaxation~\cite{parker,krstulovicbrachet2023}.

Kelvin spectroscopy thus acquires a predictive role in nonlinear vortex dynamics. In the soft-phonon limit, the parent and daughter energies fix the leading odd shear coefficient, providing an independent matching condition for effective filament theories~\cite{horn,sonin}. The daughter-momentum dependence also offers a test of how nonlocal flow enters the interaction. More broadly, this is a concrete relation between the spectrum of excitations bound to a topological defect and their coupling to the bulk Goldstone mode. In atomic gases, a comparison of initial sound phases separated by $\pi$ could probe this conversion and its ability to redistribute Kelvin occupations and change vortex geometry~\cite{bretin,parker2012}. The finite-wavelength response is governed by the independently calculated connected interaction. With an initially empty outgoing sound field, the splitting channel supplies Kelvin action while removing Kelvin energy. Incorporating this source into kinetic descriptions would allow its contribution to be assessed alongside conservative scattering and pair annihilation. Predicting the geometry left after dissipation then requires following Kelvin energy and number, together with the redistribution of mode occupations across Kelvin wave numbers.

\onecolumngrid
\clearpage
\begin{center}
\large\bfseries Supplemental Material for Kelvin-Wave Multiplication by Sound Emission
\end{center}
\twocolumngrid
\setcounter{equation}{0}
\setcounter{figure}{0}
\setcounter{table}{0}
\setcounter{section}{0}
\renewcommand{\theHequation}{S.\arabic{equation}}
\renewcommand{\theHtable}{S.\arabic{table}}
\renewcommand{\theHfigure}{S.\arabic{figure}}

\raggedbottom
\renewcommand{\theequation}{S\arabic{equation}}
\renewcommand{\thetable}{S\arabic{table}}
\renewcommand{\thefigure}{S\arabic{figure}}
\renewcommand{\thesection}{S\arabic{section}}
\setcounter{secnumdepth}{1}

\section{GP modes and Kelvin observables}

Our length, energy and time units are $\xi=\hbar/\sqrt{2m_a\mu}$, $\mu=gn_0$ and $\hbar/\mu$; $m_a,n_0,g$ are atomic mass, bulk density and contact coupling. The dimensionless GP field obeys $i\dot\psi=(-\nabla^2+|\psi|^2-1)\psi$. A vortex with winding number $+1$ has the stationary form $f(r)e^{i\varphi}$, where
\begin{equation}
 \begin{gathered}
 -f''-f'/r+f/r^2+(f^2-1)f=0,\\
 f(0)=0,\qquad f(\infty)=1.
 \end{gathered}
 \label{sm-eq:profile}
\end{equation}
Set $N_\xi=n_0\xi^3$ and $F=2\pi LN_\xi$, with axial period $L$. The factor $F$ normalizes the angular and axial mode functions. A coherent amplitude $a_j=\langle\widehat b_j\rangle/\sqrt F$ corresponds to quantum occupation $F|a_j|^2$. For $\widehat\Psi=\sqrt{n_0}e^{i\varphi}(f+\widehat\eta)$,
\begin{equation}
 \begin{aligned}
 \widehat\eta=\frac1{\sqrt F}\sum_{m,k,\lambda}\big[&u_{mk\lambda}\widehat b_{mk\lambda}e^{i(m\varphi+kz)}\\
 &-v_{mk\lambda}^*\widehat b_{mk\lambda}^\dagger e^{-i(m\varphi+kz)}\big].
 \end{aligned}
 \label{sm-eq:expansion}
\end{equation}
The integer $m$ is relative to the background winding, $k=2\pi j/L$, and $\lambda$ labels radial modes. Vortex-bound modes are localized near the vortex and have unit Bogoliubov norm, the particle--hole norm $\int r\,dr(|u|^2-|v|^2)=1$; bulk modes have $\delta(s-s')$ normalization in radial momentum $s>0$. The physical Nambu spinor $U=(u,-v)^T$ satisfies
\begin{align}
 \mathcal K_{m,k}U&=eJU,\qquad J=\operatorname{diag}(1,-1),\nonumber\\
 \mathcal K_{m,k}&=\begin{pmatrix}h_{m+1,k}&f^2\\f^2&h_{m-1,k}\end{pmatrix},\nonumber\\
 h_{\ell,k}&=-r^{-1}\partial_r r\partial_r+\ell^2/r^2+k^2+2f^2-1.
 \label{sm-eq:bdg}
\end{align}
The fundamental Kelvin branch has $m=-1$, even energy $E_K(k)$, and no radial core node. The collective translation mode is kept separate. Bulk sound has $\epsilon(p)=p\sqrt{p^2+2}$ and $c=\sqrt2$. In splitting, $q=k-k_1-k_2$, $\Omega=E_K(k)-E_K(k_1)-E_K(k_2)>0$, and $p=\sqrt{q^2+s^2}$, with radial sound momentum $s>0$.

\paragraph{Core weight and angular charge.}
Near the axis, the stationary field is $\psi_0\simeq A(x+iy)$, with $A=f'(0)$. For an $m=-1$ Kelvin mode, the particle component is finite at the axis, while regularity makes the hole component vanish as $r^2$. To first order, the field zero therefore satisfies $Aw_K+\delta\psi(0)=0$, giving
\begin{equation}
 \widehat w_K(z)=-\frac1{A\sqrt F}\sum_k u_k(0)\widehat b_ke^{ikz}.
 \label{sm-eq:core}
\end{equation}
The vacuum-subtracted mean-square displacement and mean-square slope are consequently
\begin{equation}
 \begin{gathered}
 W_K=F^{-1}\sum_k Q(k)n_k,\quad
 S_K=F^{-1}\sum_k k^2Q(k)n_k,\\
 Q(k)=|u_k(0)/A|^2.
 \end{gathered}
 \label{sm-eq:geometry}
\end{equation}
Here $n_k=\langle\widehat b_k^\dagger\widehat b_k\rangle$. Thus $Q(k)/F$ is the contribution of one Kelvin quantum to the mean-square displacement. These expressions give the Kelvin contributions to the two observables in linear theory. The instantaneous field zero also contains sound and nonlinear deformations. The translation profiles $(f'+f/r)/2$ and $(f/r-f')/2$ have Bogoliubov norm $1/2$, giving $u_k(0)\to\sqrt2 A$ and $Q(k)\to2$ as $k\to0$. Thus $W_{K,\mathrm{phys}}\simeq\mathcal N_K/(\pi n_0L_{\mathrm{phys}})$ at long wavelengths, where $\mathcal N_K=\sum_k n_k$ includes both axial directions.

The generator $L_z-\hbar\mathcal N_{\rm at}$ subtracts the angular momentum of the winding-$+1$ background. A unit-norm quasiparticle carries excess angular momentum $m\hbar\int r\,dr(|u|^2-|v|^2)=m\hbar$. At fixed atom number this is also its physical angular-momentum change, giving $-\hbar$ per Kelvin quantum. With other bound modes and collective motion fixed, conservation gives $\Delta\mathcal N_K=\mathcal J_B$, where $\mathcal J_B=\sum m n^B_{m,k,s}$ is sound angular momentum in units of $\hbar$. Other bound branches contribute $m\Delta n$; finite-radius measurements also include sound stored inside.

\section{From rotations to acoustic shear}
\label{sm-sec:shear}

We consider Kelvin wavelengths much larger than the core radius in a zero-temperature, single-component Galilean fluid. The vortex is force-free, the matched core response is regular, and no additional low-energy core resonance is present. We take long Kelvin wavelengths at fixed daughter fractions and softer sound, $p\ll\min(k_1,k_2)$. For $E_K\simeq\Lambda k^2$ with slowly varying $\Lambda$ and fixed daughter fractions $k_1\simeq\alpha_0k$, $k_2\simeq(1-\alpha_0)k$,
\[
 \frac{\Omega}{E_K(k)}\simeq2\alpha_0(1-\alpha_0),\qquad
 \frac pk\simeq\frac{2\alpha_0(1-\alpha_0)k\Lambda}{c}\to0.
\]
Thus sound can carry a finite fraction of the energy while its momentum is small relative to the Kelvin momenta. The fluid displacement $B$ obeys $v=\dot B$ and $\delta n=-\nabla\cdot B$, with $\delta n$ the fractional density perturbation. For a line $X$ and $\kappa=2\pi N_\xi$, the Berry term is first order in time and generates transverse vortex motion, while the mixing term couples it to fluid displacement~\cite{sm-horn}:
\begin{equation}
 \begin{aligned}
 S_{\rm B}&=-\frac\kappa3\int dt\,dz\,X\cdot(\dot X\times X'),\\
 S_{\rm mix}&=\kappa\int dt\,dz\,B(X)\cdot(\dot X\times X').
 \end{aligned}
 \label{sm-eq:berry}
\end{equation}
Dots and primes denote $\partial_t$ and $\partial_z$. Since $\delta S_{\rm B}=-\kappa\int dt\,dz\,\delta X\cdot(\dot X\times X')$, the coordinate change $X=Y+B(Y,t)$ cancels the leading mixing. Writing $Y=(x,y,z)$ reparametrizes the displaced line by its axial coordinate. It describes motion relative to the fluid, with gradients still deforming the line. Matching to the core includes the outer potential flow in the coefficients.

For local tension $T=\kappa\Lambda$, the remaining strain term is $-T(Y_i'Y_j'/|Y'|)\partial_iB_j$. With $w=x+iy$ and $\pi=(x,y)$, its cubic Kelvin part at leading order in sound gradients is
\begin{equation}
 L_{K^3B}=\frac T2|w'|^2\pi'_a(\partial_aB_z+\partial_zB_a).
 \label{sm-eq:strain}
\end{equation}
The sound field therefore couples through mixed transverse--axial shear. The response of the nonlocal vortex energy follows from rotational invariance and the quadratic Kelvin energy~\cite{sm-sonin},
\begin{equation}
 H_2=\frac\kappa2\int dz\,w^*E_K(-i\partial_z)w.
 \label{sm-eq:h2}
\end{equation}
In laboratory coordinates, rotation gives $(\delta x,\delta z)=(\varepsilon z,-\varepsilon x)$, whereas symmetric $xz$ shear gives $(\varepsilon z,+\varepsilon x)$; both have $\delta y=0$. Expressing the displaced line at the same laboratory axial coordinate gives
\[
 \left.\delta w\right|_z=\delta w_{\rm lab}-w'\delta z.
\]
Thus $\delta_{\rm rot}w=\varepsilon z+\varepsilon xw'$ and $\delta_{\rm sh}w=\varepsilon z-\varepsilon xw'$, with $x=(w+w^*)/2$. The tilt $\varepsilon z$ replaces one amplitude in $H_4$, leaving a cubic term. The reparametrization $xw'$ contains two amplitudes, so varying $H_2$ also gives a cubic term. Here $H_4$ is the quartic part of the effective vortex-line Hamiltonian, with the core and outer flow already matched; Sec.~\ref{sm-sec:vertex} gives the bare GP interaction.

Let $[\cdot]_{P12}$ extract the coefficient of the labeled monomial $w_kw_{k_1}^*w_{k_2}^*$, with the axial momentum factor $L\delta_{k,k_1+k_2}$ removed, at zero recoil $k=k_1+k_2$. Put $E_i=E_K(k_i)$. Fourier substitution, retaining both creation orderings, gives
\begin{equation}
 \begin{gathered}
 \bigl[\delta H_2(-\varepsilon xw')\bigr]_{P12}
 =\frac{i\varepsilon\kappa k}{4}\Delta E_0,\\
 \Delta E_0=E_K(k)-E_1-E_2.
 \end{gathered}
 \label{sm-eq:regauge}
\end{equation}
Varying $w^*$ gives $+i\varepsilon\kappa kE_K(k)/4$; varying $w$ gives $-i\varepsilon\kappa k(E_1+E_2)/4$. In the latter, the energy operator acts on products such as $w_kw_{k_1}^*$. At zero recoil this product has wave number $k-k_1=k_2$, yielding $E_K(k_2)$; exchanging daughters yields $E_K(k_1)$. A rigid rotation cannot change the energy, so the quartic tilt and quadratic reparametrization contributions in $\delta_{\rm rot}w$ cancel:
\begin{equation}
 \delta_{\rm tilt}H_4+\delta_{+\rm reg}H_2=0,\qquad
 \delta_{-\rm reg}H_2=-\delta_{+\rm reg}H_2.
 \label{sm-eq:rotation-cancellation}
\end{equation}
Here $\delta_{\pm\rm reg}w=\pm\varepsilon xw'$. The quartic tilt is the same for rotation and shear, while the coordinate-correction sign reverses. Therefore $\delta_{\rm sh}H=\delta_{\rm tilt}H_4+\delta_{-\rm reg}H_2=2\delta_{-\rm reg}H_2$ at this order, giving
\begin{equation}
 [\delta_{\rm sh}H]_{P12}=\frac{i\varepsilon\kappa k}{2}\Delta E_0.
 \label{sm-eq:ward}
\end{equation}
The rotationally invariant canonical filament Hamiltonian includes core matching and, for a fixed axial cutoff, the corresponding cutoff-matching terms. For longitudinal sound, $\partial_aB_z=\partial_zB_a$, so Eq.~\eqref{sm-eq:strain} is exactly the symmetric shear of Eq.~\eqref{sm-eq:ward}. Its factor $\sin\theta\cos\theta e^{i\varphi}$ has $\ell=2,m=1$; $\theta$ is the polar angle from the axis and $\varphi$ the transverse azimuth.

The translation normalization is $w=-\sqrt{2/F}\sum b_ke^{ikz}$. The quadratic sound action $N_\xi\int dt\,d^3x[\dot B^2-c^2(\nabla\cdot B)^2]/4$ gives radial displacement coefficient $\sqrt{s}/\sqrt{F\Omega}$. We choose the $m=+1$ sound-mode phase so that the coefficient of the far-field density Bessel function is real and positive. Then
\begin{equation}
 \partial_+B_z^*(0)=-i\frac qp\frac{s^{3/2}}{\sqrt{F\Omega}},\qquad
 \partial_+=\partial_x+i\partial_y.
 \label{sm-eq:soundnorm}
\end{equation}
The factor $q/p$ selects axial longitudinal displacement. Near the axis, $J_1(sr)e^{-i\varphi}\simeq s(x-iy)/2$, so $\partial_+$ supplies one power of $s$. Radial continuum normalization supplies the remaining $\sqrt{s}$, and oscillator normalization gives $1/\sqrt\Omega$. For general transverse shear, $\varepsilon$ becomes $\varepsilon_x+i\varepsilon_y$. For the created phonon, this shear component is $\partial_+B_z^*(0)$, where $B_z^*$ is the normalized mode function multiplying the phonon creation operator. Restoring the axial integral and the three Kelvin external legs gives
\[
 \begin{aligned}
 \frac{\mathcal M_{\rm sh}}F
 &=\frac{i\kappa Lk\Delta E_0}{2}\left(-\sqrt{\frac2F}\right)^3\\
 &\quad\times\left[-i\frac qp\frac{s^{3/2}}{\sqrt{F\Omega}}\right].
 \end{aligned}
\]
Using $\kappa L=F$ yields
\begin{equation}
 \mathcal M_{\rm sh}=-\sqrt2 k\Delta E_0\frac qp\frac{s^{3/2}}{\sqrt\Omega},
 \label{sm-eq:shear}
\end{equation}
with physical labeled amplitude $\mathcal M_{\rm sh}/F$.

\paragraph{Scope and finite-wave-number comparison.}
A static divergence-free displacement changes neither density nor velocity; its coordinate deformation is already included. Independent density-dependent tension and relative-velocity couplings remain allowed~\cite{sm-horn}. The $m=+1$ density perturbation vanishes on axis, so evaluating it on the displaced line adds a transverse gradient. Its leading term is even in $q$; an odd recoil correction brings $p/k$ at fixed daughter fractions. After uniform motion is removed, additional time derivatives suppress the inertial contribution by $[E_K(k)/(ck)]^2$, the square of the Kelvin phase velocity in units of the sound speed. Thus Eq.~\eqref{sm-eq:shear} fixes the leading odd soft-sound response, with material-dependent finite-$k$ corrections.

For $E_K(k)=k^2[\ln(1/k)+C]$ and $k_1=\alpha_0k$ at zero recoil,
\begin{equation}
 \begin{aligned}
 \frac{\Delta E_0}{2\alpha_0(1-\alpha_0)k^2}
 &=\ln(1/k)+C+h(\alpha_0),\\
 h(\alpha_0)&=\frac{\alpha_0^2\ln\alpha_0+(1-\alpha_0)^2\ln(1-\alpha_0)}
 {2\alpha_0(1-\alpha_0)}.
 \end{aligned}
 \label{sm-eq:shape}
\end{equation}
Parent-only tension $E_K(k)/k^2$ omits $h(\alpha_0)$.

\paragraph{Rate kernel and validity near soft daughters.}
After deriving the rate at fixed daughter fractions, we check its validity when one daughter momentum vanishes. Write $\alpha=k_1/(k_1+k_2)$, $\beta=1-\alpha$, and $y=q/p$; $\alpha\to\alpha_0$ as $p/k\to0$. At leading order $\Omega=2\alpha\beta k^2\Lambda$, $p=\Omega/c$, $s=p\sqrt{1-y^2}$, $dk_1dk_2\simeq kp\,d\alpha\,dy$, and $v_\perp\simeq c\sqrt{1-y^2}$. The amplitude scales as $\mathcal M_{\rm sh}\sim k^5\Lambda^2$ and the phase space as $k^3\Lambda$, giving $\Gamma_{13}\sim k^{13}\Lambda^5$. Equations~\eqref{sm-eq:width} and \eqref{sm-eq:shear} give
\begin{align}
 \frac{d\Gamma_{13}}{d\alpha\,dy}&\simeq
 \frac{4k^{13}\Lambda^5}{\pi^3c^5N_\xi^2}
 [\alpha(1-\alpha)]^5y^2(1-y^2),\nonumber\\
 \Gamma_{13}&\simeq\frac{4k^{13}\Lambda^5}{10395\pi^3c^5N_\xi^2}.
 \label{sm-eq:kernel}
\end{align}
The normalized marginals are $P(\alpha)=\alpha^5(1-\alpha)^5/\mathrm B(6,6)$ and $P(y)=15y^2(1-y^2)/4$. Averaging $\Omega/E_K(k)\simeq2\alpha(1-\alpha)$ with $P(\alpha)$ gives $6/13$; the parent's mean-square-slope contribution has the same mean fractional reduction. In the translation limit, the two daughters contribute twice the parent's mean-square displacement. The polar density follows the azimuth-integrated $|Y_{2,1}|^2$ pattern after including the final-state density and energy-shell integral.

The endpoint hierarchy follows directly from kinematics. Put $m_*=\min(k_1,k_2)$, $M_*=\max(k_1,k_2)$, $\delta=k-M_*$, and $r_*=E_K'(k)/c<1$. Convexity gives $\Omega\le E_K'(k)\delta$ and sound costs $\Omega\ge c|\delta-m_*|$, hence
\begin{equation}
 \frac{m_*}{1+r_*}\le\delta\le\frac{m_*}{1-r_*},\qquad
 \frac p{m_*}\le\frac{r_*}{1-r_*}\longrightarrow0.
 \label{sm-eq:endpoint}
\end{equation}
The normalized soft Kelvin core is finite, $Q\to2$. A zero-momentum Kelvin mode translates the vortex. An undifferentiated soft leg therefore moves the acoustic evaluation point and introduces an acoustic momentum factor $p$. A differentiated leg instead supplies its Kelvin momentum $m_*$. Together with the strain gradient, these terms give a source of order $p(m_*+p)$, with no inverse soft momentum. Equation~\eqref{sm-eq:endpoint} keeps it of order $m_*^2$. Below-sound $m=0$ responses give logarithms rather than poles; after translation cancellation, the remaining endpoint logarithms multiply the integrable $\alpha^5$ weight.

To restore physical units, let $k_{\rm phys},s_{\rm phys}$ be dimensional wave numbers and $\omega$ the physical sound frequency. For mass density $\rho_s$, circulation $\kappa_{\rm v}$, and sound speed $c_s$, the displacement normalizations are $w_{\rm phys}=-\sqrt{2\hbar/(\rho_s\kappa_{\rm v}L_{\rm phys})}\sum b_ke^{ik_{\rm phys}z_{\rm phys}}$ and $\mathcal B_{s_{\rm phys}}=\sqrt{\hbar s_{\rm phys}/(4\pi\rho_sL_{\rm phys}\omega)}$. They give
\begin{equation}
 \Gamma_{13,\mathrm{phys}}\simeq
 \frac{\hbar^2\kappa_{\rm v}^4}{2661120\pi^7\rho_s^2c_s^5}k_{\rm phys}^{13}\Lambda^5.
 \label{sm-eq:physicalrate}
\end{equation}
For a GP fluid, $\rho_s=m_an_0$ and $\kappa_{\rm v}=2\pi\hbar/m_a$ recover Eq.~\eqref{sm-eq:kernel} times $\mu/\hbar$. In helium $\rho_s$ is the full zero-temperature mass density, and the expression applies in the same thin-filament acoustic limit.

\section{Connected four-wave coefficient}
\label{sm-sec:vertex}

\paragraph{Why subsonic decay leaves two Kelvin quanta.}
Bulk sound has $\epsilon(p)=p\sqrt{p^2+2}\ge cp$, $c=\sqrt2$. Suppose $E_K'(k')<c$ for $0<k'\le k$. Radiation leaving no Kelvin quantum costs at least $ck>E_K(k)$. Leaving one at $k'$ requires $|k'|<k$ and releases
$E_K(k)-E_K(k')<c(k-|k'|)\le c|k-k'|$,
below the sound cost. The same inequalities hold for any number of sound-like final carriers, since their total energy is at least $c$ times their net axial momentum. Thus an isolated energy-losing decay must leave at least two nonzero Kelvin excitations under this carrier assumption.

For $K_k\to K_{k_1}+K_{k_2}+B_{+1}$, a negative daughter $k_1=-a$ would require $q=k-k_2+a$ but $\Omega=E_K(k)-E_K(a)-E_K(k_2)<c(k-k_2)<cq$. Both daughters therefore remain co-propagating. Two such inputs cannot annihilate into one sound quantum; opposite inputs can, emitting $m=-2$ sound.

The cubic and quartic parts of the GP Hamiltonian are
\begin{align}
 H_3/\mu&=N_\xi\int d^3x\,f(\widehat\eta^{\dagger2}\widehat\eta+
 \widehat\eta^\dagger\widehat\eta^2),\nonumber\\
 H_4/\mu&=\frac{N_\xi}{2}\int d^3x\,\widehat\eta^{\dagger2}\widehat\eta^2.
 \label{sm-eq:h34}
\end{align}
Use the incoming mode and the conjugates of outgoing modes. Conjugation reverses $(\omega,m,k)$ in the coefficient calculation. Thus take $A_0=U_k$, $A_1=\mathsf XU_{k_1}^*$, $A_2=\mathsf XU_{k_2}^*$, and $A_3=\mathsf XU_B^*$, with $\mathsf X$ exchanging Nambu components. Their frequency and momentum labels $(\omega,m,k)$ are
\begin{equation}
 \begin{gathered}
 (E_K(k),-1,k),\quad(-E_1,+1,-k_1),\\
 (-E_2,+1,-k_2),\quad(-\Omega,-1,-q),
 \end{gathered}
 \label{sm-eq:legs}
\end{equation}
where $q=k-k_1-k_2$, $\Omega=E_K(k)-E_1-E_2$, and $E_i=E_K(k_i)$. Complex conjugation is performed before contraction. For $A_i=(x_i,y_i)^T$, define
\begin{equation}
 S(A_i,A_j)=2f\binom{x_ix_j+x_iy_j+y_ix_j}
 {y_iy_j+x_iy_j+y_ix_j}.
 \label{sm-eq:source}
\end{equation}
The pair labels are $\omega_{ij}=\omega_i+\omega_j$, $m_{ij}=m_i+m_j$ and $k_{ij}=k_i+k_j$. Each pair drives a linear response:
\[
 [\omega_{ij}J-\mathcal K_{m_{ij},k_{ij}}]\chi_{ij}=S(A_i,A_j).
\]
The inverse linear-response operator (resolvent) gives the intermediate virtual field, which couples to the complementary pair. The connected labeled coefficient is:
\begin{align}
 \chi_{ij}&=[\omega_{ij}J-\mathcal K_{m_{ij},k_{ij}}]^{-1}S(A_i,A_j),\nonumber\\
 \mathcal M_{13}&=\mathcal M_4+
 \sum_{(ij|\ell n)}\int r\,dr\,S(A_\ell,A_n)^T\mathsf X\chi_{ij},\nonumber\\
 \mathcal M_4&=2\int r\,dr\sum_{\{i,j\}}x_ix_jy_\ell y_n.
 \label{sm-eq:coefficient}
\end{align}
The contact sum has six unordered choices of $\{i,j\}$ and complementary $\{\ell,n\}$. The three exchanges are $(01|23)$, $(02|13)$, and $(03|12)$; the contraction uses the ordinary transpose. With continuum sound normalization, $\mathcal M_{13}$ has physical coefficient $\mathcal M_{13}/F$, with leading odd soft-sound limit $\mathcal M_{\rm sh}$. It describes outgoing sound in an open radial domain. The finite-$k$ cavity coefficient $M_{\rm c}$ uses the discrete standing sound mode specified in Sec.~\ref{sm-sec:free}.

The quartic interaction couples the four modes directly; each exchange connects two cubic vertices through $\chi_{ij}$. Both mechanisms occur at the same order: inserting $\eta=\varepsilon\eta_1+\varepsilon^2\eta_2+\cdots$ into GP gives a direct cubic source and a source mixing $\eta_1$ with $\eta_2$. All amplitudes must therefore be added before squaring. Direct-source projection agrees with Eq.~\eqref{sm-eq:coefficient} on the tested energy shells; the algebra also reproduces the on-shell cancellation in the integrable one-dimensional contact gas~\cite{sm-ristivojevic}. Here intermediate pairs lie below bulk radiation thresholds, and $m=0$ responses below the resolved varicose branch. These intermediate deformations are nonradiative, while the final sound satisfies the outgoing-wave condition. A resonant intermediate mode requires an outgoing resolvent and separate sequential-decay treatment.

\paragraph{Energy conservation and identical daughters.}
Radially propagating final sound requires
\begin{equation}
 \Omega>\epsilon(|q|),\quad p^2=\sqrt{1+\Omega^2}-1,
 \quad s^2=p^2-q^2>0.
 \label{sm-eq:shell}
\end{equation}
The radial energy delta function gives the final-state density
\[
 \begin{gathered}
 \int_0^\infty ds\,\delta\!\left[\Omega-\epsilon\!\left(\sqrt{q^2+s^2}\right)\right]
 =\frac1{v_\perp},\\
 v_\perp=\epsilon'(p)\frac sp.
 \end{gathered}
\]
Integration over the remaining energy-conserving final states gives the spontaneous decay rate
\begin{equation}
 \Gamma_{13}=\frac1{16\pi^3N_\xi^2}\int dk_1dk_2\,
 \frac{|\mathcal M_{13}|^2}{v_\perp}.
 \label{sm-eq:width}
\end{equation}
The prefactor includes $1/2$ for daughter permutations. With continuum-normalized sound and both daughters in one Kelvin mode, the monomial coefficient is $\mathcal M_{13}/(2F)$. Creating two identical daughter quanta gives the Bose factor $(n_D+1)(n_D+2)$ multiplying the squared coefficient. Energy-loss events have weight $\Omega$; changes of $W_K,S_K$ have weights
\[
 \begin{aligned}
 \Delta W_K&=\frac{Q(k_1)+Q(k_2)-Q(k)}F,\\
 \Delta S_K&=\frac{k_1^2Q(k_1)+k_2^2Q(k_2)-k^2Q(k)}F.
 \end{aligned}
\]
Physical rates include $\mu/\hbar$.

\paragraph{Spontaneous decay in a finite axial period.}
For an initially empty outgoing sound continuum, take a parent at $k_P=0.60$, $L=50\pi$ and daughter momenta $k_i=0.04i$, $i=1,\ldots,15$. Of the 120 unordered pairs, 40 satisfy Eq.~\eqref{sm-eq:shell}, including four with a $0.04$ daughter. The rate per parent is
\begin{equation}
 \Gamma_{P,\mathrm{vac}}=\sum_{i\le j,\,\mathrm{open}}
 \frac{2\pi|\mathcal M_{ij}|^2}{F^2(1+\delta_{ij})v_{\perp,ij}},
 \label{sm-eq:vacuum-reference}
\end{equation}
where $\delta_{ij}$ is the Kronecker delta and $\mathcal M_{ij}$ uses Eq.~\eqref{sm-eq:coefficient}. The bound modes and scattering states are computed on radial domains of radii $400$ and $800$, respectively. Meshes $0.02,0.01,0.005$ give $N_\xi^2\Gamma_{P,\mathrm{vac}}=(1.17373,1.17176,1.17127)\times10^{-8}$. Quadratic extrapolation gives $1.17110284\times10^{-8}$, or $2.93\times10^{-11}\mu/\hbar$ at $N_\xi=20$.

\paragraph{Microscopic shear comparison.}
In the microscopic comparison, energy-conserving configurations at $y=q/p=\pm0.5$ include recoil, use $k_1=\alpha_0k$, and define
\begin{equation}
 \mathcal C(y)=-\frac{\mathcal M_{13}\sqrt\Omega}
 {2\sqrt2 k k_1k_2 y s^{3/2}},\quad
 \Lambda_{\rm odd}^{\rm GP}=\tfrac12[\mathcal C(0.5)+\mathcal C(-0.5)].
 \label{sm-eq:reduced}
\end{equation}
Dividing by $y=q/p$ makes the odd amplitude even in $\mathcal C(y)$, so averaging extracts the leading odd shear term as recoil vanishes. Finite wave numbers leave unequal sound energies and recoil corrections. For $\alpha_0=0.25,0.4,0.5$, discrepancies from the dispersion value are $(2.872,2.837,2.727)\%$ at $k=0.2$ and $(0.866,0.737,0.665)\%$ at $0.1$; at $k=0.05$, $\alpha_0=0.4$ gives $0.166\%$. Amplitudes and kinematic quantities are quadratically extrapolated as $(4X_{h/2}-X_h)/3$, with mesh pairs $0.005/0.0025$ for $0.2,0.1$ and $0.0025/0.00125$ for $0.05$. At $0.05$, paired $|y|=0.25,0.75,0.90$ give $-0.011\%,0.432\%,0.582\%$. The smaller even source permits residual radiation at $q=0$.

\begin{figure}[t]
\includegraphics[width=\columnwidth]{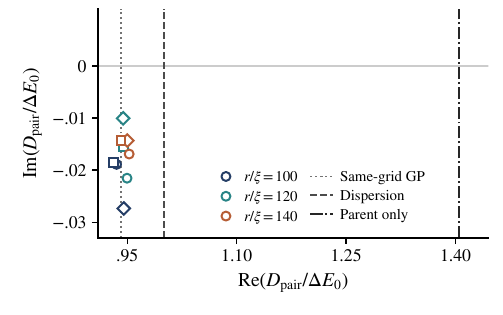}
\caption{Paired complex source at $k_P=0.35$. Colors denote $r=100,120,140$; circles, diamonds and squares use the final two, three and four carrier periods. All nine extractions are shown. Lines give the real same-grid vertex, leading dispersion prediction and parent-only estimate; their imaginary references are zero. The spread reflects the chosen extraction radii and averaging windows.}
\label{sm-fig:paired}
\end{figure}

\paragraph{Paired complex-source check.}
The Kelvin seeds drive sound at $\Omega=E_P-2E_D$, where $E_P=E_K(k_P)$ and $E_D=E_K(k_D)$, with leading field proportional to $a_P(a_D^*)^2$. Demodulation, division by this seed product and outgoing-wave calibration against a unit source extract the coupling in the normalization of Eq.~\eqref{sm-eq:coefficient}. This check uses two unforced trajectories at $k_P=0.35$, with repeated daughters $7k_P/15$ or $8k_P/15$ and $q_\pm=\pm k_P/15$. Both seed magnitudes are $0.08$, with phases $0.37,-0.51$. On the invariant lattice $q_{mn}=-mk_D+n(k_P-k_D)$, the same-grid vortex and modes use $R=400$, $h=0.04$, $|m|,|n|\le4$, $\Delta t=0.04$, $T=600$, axial period $L=30\pi/k_P$, and a quartic absorber outside $200$ reaching $0.05$ at $400$. For the target $Z=(\eta_{1,1},\eta_{-1,-1}^*)^T$, divide $\langle e^{i\Omega t}Z\rangle$ by $a_P(0)[a_D^*(0)]^2$ to obtain $\widehat Z$. With regular scattering mode $U_B$ and outgoing unit-source wave $Z_{\rm unit}$,
\begin{equation}
 \mathcal M_{\rm dyn}=2\frac{\mathcal W_r(U_B,\widehat Z)}{\mathcal W_r(U_B,Z_{\rm unit})},\quad
 \mathcal W_r(U,Z)=r(U^\dagger Z'-U'^\dagger Z).
 \label{sm-eq:dynamic}
\end{equation}
The Wronskian uses the ordinary two-component inner product; the factor two restores repeated-daughter labeling. Defining $\mathcal V_\pm=\mathcal M_\pm^{\rm dyn}\sqrt{\Omega_\pm}/s_\pm^{3/2}$ and $y_\pm=q_\pm/p_\pm$ gives
\begin{equation}
 D_{\rm pair}=-\frac{\mathcal V_+-\mathcal V_-}{\sqrt2 k_P(y_+-y_-)},
 \label{sm-eq:pair}
\end{equation}
whose shear value is $\Delta E_0=E_K(k_P)-2E_K(k_P/2)$. Figure~\ref{sm-fig:paired} gives real ratios $0.9307$--$0.9527$, versus $0.9418$ from the microscopic amplitude evaluated on the same spatial grid and $1.40465$ from parent-only tension. The imaginary part of the ratio ranges from $-0.0274$ to $-0.0100$. Let $z_{r,n_c}=D_{\rm pair}^{\rm dyn}/\Delta E_0$ label the nine extractions and $z_{\rm GP}=D_{\rm pair}^{\rm GP}/\Delta E_0\simeq0.9418$ the same-grid vertex. The largest relative complex difference from that vertex is
\[
 \epsilon_{\rm src}=\max_{r,n_c}
       \frac{|z_{r,n_c}-z_{\rm GP}|}{|z_{\rm GP}|}=2.93\%.
\]
The denominator is the microscopic amplitude's modulus; the leading dispersion prediction is one. Averages use $[600-2\pi n_c/\Omega_\pm,600]$, $n_c=2,3,4$. Fit the measured field and radial derivative to incoming/outgoing waves and project onto $e^{-i\Omega_\pm t}$. Over both branches and all nine radius/window choices, $\max|C_{\rm in}/C_{\rm out}|=5.92\%$. The minus-branch $r=140$ four-period window starts $0.052$ periods after group arrival, leaving little time for the arrival transient to decay. The comparison uses the listed seed amplitudes, time step and Fourier cutoff at finite $k$, retaining separate recoil kinematics on each branch.

An interaction check uses $H=\int d^3x[|\nabla\psi|^2+(n-1)^2/2+\ell_\rho^2|\nabla n|^2/2]$, $n=|\psi|^2$, with $\ell_\rho^2=0.5$~\cite{sm-fu2003,sm-collin2007}. Its bulk dispersion is $\Omega_\rho^2=2p^2+(1+2\ell_\rho^2)p^4$. For each model, the two daughter wave numbers are equal, and the $q/p=\pm0.5$ shells are retuned separately. Recomputing the core, modes, contact and exchange terms, and current matrix $G_\rho=I_2+\ell_\rho^2f^2ee^T$, $e=(1,1)^T$, gives the paired source $D_{\rm pair}/\Delta E_0=0.9832$ after two-mesh extrapolation at $k=0.3$; the contact model gives $0.9863$. This modification tests the sensitivity of the relation to vortex-core structure within a specified GP Hamiltonian.

\section{Coherent conversion in GP evolution}
\label{sm-sec:free}

The vortex solves Eq.~\eqref{sm-eq:profile} with $f(R)=0$; all modes and responses obey this reflecting wall. Table~\ref{sm-tab:initial} lists the parent $P$, daughter $D$ and discrete sound mode $B$, each normalized to unit positive Bogoliubov norm. For $L=40\pi$, their invariant Fourier lattice is
\begin{equation}
 q_{m\nu}=-0.25m+0.35\nu,\qquad m,\nu\in\mathbb Z.
 \label{sm-eq:screw}
\end{equation}
It is closed under the local GP nonlinearity and excludes a uniform $|m|=1$, $q=0$ core translation. The daughter index $(m,\nu)=(-1,0)$ therefore corresponds to axial momentum $q_D=0.25$.

\paragraph{Preparation and three-mode Hamiltonian.}

Write $a_j=\langle\widehat b_j\rangle/\sqrt F$ and $I_j=|a_j|^2$. The coherent occupation is $FI_j$, whereas $n_k$ in Sec.~S1 is absolute occupation. With $\phi_B^{(0)}=\phi_B(0)$ from Table~\ref{sm-tab:initial}, the opposite preparation adds $\pi$ and its consistent quadratic response. Initial conserved quantities agree to roundoff.

\begin{table}[t]
\caption{Initial cavity modes for the amplifying preparation, with $R=13.3177377165$ and $L=40\pi$. The opposite phase uses $\phi_B^{(0)}+\pi$.}
\label{sm-tab:initial}
\begin{ruledtabular}\begin{tabular}{lrrr}
Quantity & Parent $P$ & Daughter $D$ & Sound $B$\\
$m$ & $-1$ & $-1$ & $+1$\\
$q$ & $0.60$ & $0.25$ & $0.10$\\
$E_j$ & $0.46593605$ & $0.10684666$ & $0.25460068$\\
$|a_j(0)|$ & $0.24$ & $0.12$ & $0.20$\\
$\phi_j(0)$ & $0.37$ & $-0.51$ & $2.9607963268$
\end{tabular}\end{ruledtabular}
\end{table}

Alongside the three linear modes, the initial field includes their second-order GP response: one for each of the $6(6+1)/2=21$ unordered pairs formed from the modes and their conjugates, including $B+B$. Static density responses have a fixed condensate phase. At fixed seed amplitudes, a response along $f_\mu$ enforces the total atom number through a scalar norm constraint; $\mathcal L_+f_\mu=f$, with $\mathcal L_+=-r^{-1}\partial_r r\partial_r+r^{-2}+3f^2-1$.

The $P,D,B$ Hamiltonian per $\mu F$, calculated microscopically before GP evolution, is
\begin{equation}
 \begin{aligned}
 h_{\rm c}={}&\sum_{j=P,D,B}E_j I_j+\tfrac12\sum_{i,j}\lambda^N_{ij}I_i I_j\\
 &+\tfrac{M_{\rm c}}2(a_Pa_D^{*2}a_B^*+\mathrm{c.c.}),
 \qquad I_j=|a_j|^2,
 \end{aligned}
 \label{sm-eq:cavity-h4}
\end{equation}
where $M_{\rm c}=-0.0103795933$ is the full connected coefficient for unit-normalized discrete modes, and
\begin{equation}
 10^3\lambda^N=\begin{pmatrix}
 4.91305&-21.59324&0.235085\\
 -21.59324&-0.0762104&-0.238680\\
 0.235085&-0.238680&3.05595
 \end{pmatrix}.
 \label{sm-eq:cavity-shifts}
\end{equation}
The shifts $\lambda^\mu$ hold the chemical potential fixed; $\lambda^N$ instead hold total atom number fixed. They satisfy $\lambda^N_{ij}=\lambda^\mu_{ij}+d_i d_j/(2\chi)$, where $d_j$ is the change in $\mathcal N_{\rm at}/F$ per unit $I_j$, including its static second-order response, and $\chi=\int r\,dr\,ff_\mu$. The static response $\zeta_j$ to a mode and its conjugate gives
\[
 d_j=\int r\,dr\,[|u_j|^2+|v_j|^2+2f\zeta_j],
 \qquad \zeta_j=(\chi_{j\bar j})_1,
\]
where $\bar j$ denotes the conjugate leg. We obtain $(d_P,d_D,d_B)=(-0.06661285,-0.02158680,-0.11972318)$ and $\chi=39.36922$. To quadratic order the compensating background shift satisfies
\[
 \begin{gathered}
 \delta(\mathcal N_{\rm at}/F)=\sum_jd_j I_j+2\chi\,\delta\mu=0,\\
 \delta\mu=-\frac{\sum_jd_jI_j}{2\chi}.
 \end{gathered}
\]
Eliminating this shift gives the correction to $\lambda^N$. We tune $R$ to initial resonance, $\widetilde E_P=2\widetilde E_D+\widetilde E_B$ to preparation accuracy, where $\widetilde E_j=E_j+\sum_i\lambda^N_{ji}I_i$. Changing occupations can detune it. 

\paragraph{Conversion and initial sound phase.}
Hamilton's equations for Eq.~\eqref{sm-eq:cavity-h4} give
\[
 \begin{gathered}
 (\dot I_P,\dot I_D,\dot I_B)=(-1,2,1)J_{\rm c},\\
 J_{\rm c}=M_{\rm c}I_D\sqrt{I_PI_B}\sin\vartheta,
 \end{gathered}
\]
where $\vartheta=\phi_P-2\phi_D-\phi_B$ and $\phi_j=\arg a_j$. Each event removes one parent and creates two daughters and one phonon. Thus $K_3=I_P+I_D$ changes at the rate $\dot I_B$. Shifting the sound phase by $\pi$ reverses the initial rate. The subsequent evolution of the occupations and phases changes the transfer rate in each trajectory.

\paragraph{Comparison with GP evolution.}
All three model amplitudes evolve freely. GP evolution uses the field equation directly, while $M_{\rm c}$ and $\lambda^N$ determine the separate three-mode prediction. At saved GP times $t_\ell=0.1\ell$, $\ell=0,\ldots,10000$, use $d_{\rm GP}=[a_D^+-a_D^-]e^{iE_Dt_\ell}/2$ after removing the condensate phase as defined below. The exponential removes linear oscillation. Apply the same operation to the model and interpolate its real and imaginary parts linearly onto these times, giving $d_{\rm model}$. The relative discrepancy is
\[
 \epsilon_D=\left[\frac{\sum_{\ell=0}^{10000}|d_{\rm GP}(t_\ell)-d_{\rm model}(t_\ell)|^2}
 {\sum_{\ell=0}^{10000}|d_{\rm model}(t_\ell)|^2}\right]^{1/2}=3.90\%.
\]
All samples have equal weight, and the model and GP signals are compared directly in the common amplitude and phase convention. The terminal GP/prediction ratio is $1.04184-0.00200i$. With $M_{\rm c}=0$ and shifts retained, the phase-odd response vanishes. The bulk dispersion at $E_B$ gives $p/k_D\sim0.7$, so this comparison probes the full finite-$k$ coupling.

\paragraph{Average over the initial sound phase.}
Set $I_{j0}=I_j(0)$. A uniform initial $\phi_B$ gives $\langle\dot K_3(0)\rangle_{\phi_B}=0$ and
\begin{equation}
 \begin{aligned}
 \langle\ddot K_3(0)\rangle_{\phi_B}
 &=\frac{M_{\rm c}^2}{2}[4I_{P0}I_{D0}I_{B0}\\
 &\qquad+I_{D0}^2(I_{P0}-I_{B0})]>0
 \end{aligned}
 \label{sm-eq:phase-curvature}
\end{equation}
For the stated initial occupations, this positive curvature gives an initial phase-averaged gain proportional to $t^2$. We test the finite-time mean with eight GP preparations, $\phi_B^{(j)}=\phi_B^{(0)}+\pi j/4$, $j=0,\ldots,7$, at the same $R,L$, amplitudes and $T=1000$. Each has a phase-consistent quadratic response and fixed atom number; $j=0,4$ are the preceding pair. For each preparation, $K_{\rm b}$, defined below, sums all nine retained Kelvin modes. We then average over the eight initial sound phases and divide by the mean initial value. Initial energy and momenta vary with phase; the excess grand-energy range is $0.2227\%$ of its mean. Each trajectory conserves these quantities. The mean retained Kelvin occupation grows by $5.77\%$, while the phase-averaged bound in Eq.~\eqref{sm-eq:band-bound} guarantees a gain of at least $3.04\%$. The four-node subset $j=0,2,4,6$ changes the terminal gain by $0.0001407$ percentage points. 

\paragraph{Kelvin-mode and vortex-position measurements.}

To include gain outside the selected modes, we project GP onto all retained fundamental Kelvin modes. Remove the uniform condensate phase $\phi$, obtained from the overlap of the $m=\nu=0$ field with $f$, by $\widetilde\eta=e^{-i\phi}(f+\eta)-f$. The symplectic projections are
\begin{equation}
 \begin{gathered}
 a_\nu=\int r\,dr\,[u_\nu^*\widetilde\eta_{-1,\nu}
 +v_\nu^*\widetilde\eta_{+1,-\nu}^*],\\
 K_{\rm b}=\sum_\nu|a_\nu|^2,\qquad
 E_{\rm b}=\sum_\nu E_K(q_\nu)|a_\nu|^2,
 \end{gathered}
 \label{sm-eq:band}
\end{equation}
Here $q_\nu=0.25+0.35\nu$ includes both directions, with all nine $\nu=-4,\ldots,4$ retained at baseline. Thus $K_{\rm b}$ and $E_{\rm b}$ are the occupation per $F$ and quadratic energy per $\mu F$ of this specified linear Bogoliubov basis. The selected sound occupation is $N_B=|a_B|^2$.

We determine the instantaneous vortex-core position from $\psi[X(z),Y(z),z]=0$, set $w_{\rm core}=X+iY$, and define
\begin{equation}
 \begin{gathered}
 W_{\rm core}=L^{-1}\int_0^L|w_{\rm core}|^2dz,\\
 S_{\rm core}=L^{-1}\int_0^L|\partial_z w_{\rm core}|^2dz,\\
 w_D=L^{-1}\int_0^Lw_{\rm core}(z)e^{-ik_Dz}dz.
 \end{gathered}
 \label{sm-eq:raw-core}
\end{equation}
Fit each angular field by unweighted least squares to $f\delta_{m0}\delta_{\nu0}+\eta_{m\nu}\simeq r^{|m+1|}\sum_{j=0}^{d}c_jr^{2j}$ near the axis; complex Newton iteration locates the zero at 512 physical axial points. Table~\ref{sm-tab:coherent} uses $r<0.8\xi$, $d=3$. The four controls use $r<0.6\xi$ or $0.8\xi$ and $d=2$ or $3$.

\begin{table}[t]
\caption{GP observable ratios $X(T)/X(0)$ at $T=1000$ for the two initial sound phases. Core values use $r<0.8\xi$, $d=3$.}
\label{sm-tab:coherent}
\begin{ruledtabular}\begin{tabular}{lrr}
Observable ratio & $\phi_B^{(0)}$ & $\phi_B^{(0)}+\pi$\\
$K_{\rm b}/K_{\rm b}(0)$ & $1.17573$ & $0.93922$\\
$E_{\rm b}/E_{\rm b}(0)$ & $0.88073$ & $1.04273$\\
$N_B/N_B(0)$ & $1.31092$ & $0.89161$\\
$W_{\rm core}/W_{\rm core}(0)$ & $1.10482$ & $0.91837$\\
$S_{\rm core}/S_{\rm core}(0)$ & $0.83041$ & $1.02019$\\
$|w_D|/|w_D(0)|$ & $1.62753$ & $0.59294$
\end{tabular}\end{ruledtabular}
\end{table}
The four plus fits give $W_{\rm core}(T)/W_{\rm core}(0)=1.10478$--$1.10486$ and $S_{\rm core}(T)/S_{\rm core}(0)=0.83037$--$0.83044$. Using $t=200$ as the reference, the changes $100[X(1000)/X(200)-1]$ for plus are $+15.04\%$ in $K_{\rm b}$, $-10.50\%$ in $E_{\rm b}$, $+13.23\%$ in $W_{\rm core}$ and $-11.93\%$ in $S_{\rm core}$; minus reverses these signs. The leading initial mean-square Kelvin slope is $0.05215$, while the actual core value is $0.05410$.

\paragraph{Angular-momentum balance and unassigned field.}
To retain angular momentum outside the projected modes, write $\widetilde\eta=\eta_K+\eta_B+R_\perp$, with the first two fields reconstructed from the linear projections. With the scalar inner product $\langle X,Y\rangle=\sum_{m,\nu}\int r\,dr\,X_{m\nu}^*Y_{m\nu}$, define $\mathcal J_R=\langle R_\perp,mR_\perp\rangle$ and $C_R=2\operatorname{Re}\langle\eta_K+\eta_B,mR_\perp\rangle$. The exact field identity is
\begin{equation}
 \begin{aligned}
 \mathcal J&\equiv\frac{L_z/\hbar-\mathcal N_{\rm at}}F\\
 &=-K_{\rm b}+N_B+\mathcal J_R+C_R.
 \end{aligned}
 \label{sm-eq:charge}
\end{equation}
Here $\mathcal N_{\rm at}$ is total atom number. Orthogonality gives $|C_R|<3.2\times10^{-17}$, retained explicitly. The mode projection keeps the negative-$q$ Kelvin contribution to $K_{\rm b}$ separate from the sound contribution in the same angular pair.

The directly measured $K_{\rm b}$ rises from $0.0720000$ to $0.0846528$ for plus. Independently, set $B_R=\langle R_\perp,|m|R_\perp\rangle$, so $|\mathcal J_R|\le B_R$. Equation~\eqref{sm-eq:charge} implies
\begin{equation}
 \begin{aligned}
 \Delta K_{\rm b}\ge{}&\Delta N_B-B_R(0)-B_R(T)\\
 &-|\Delta\mathcal J|-|\Delta C_R|.
 \end{aligned}
 \label{sm-eq:band-bound}
\end{equation}
For $T=1000$, $\Delta N_B=0.01243687$ and the two endpoint bounds are $0.0008238654$ and $0.0014950638$. This guarantees an increase exceeding $14.05\%$ of $K_{\rm b}(0)$ for the worst allowed remainder charge. The signed remainder charge changes by $0.0002159512$.

\paragraph{Quadratic reconstruction of the selected three modes.}
For a separate check, reconstruct the selected $P,D,B$ linear field $X_1$ and its quadratic and fixed-number correction $\chi_2$ in the normalization and inner product of $\widetilde\eta$. Here $X_1$ is the three-mode part of $\eta_K+\eta_B$. With $R_2=\widetilde\eta-X_1-\chi_2$ and $\mathcal J[X]=\langle X,mX\rangle$, the omitted angular momentum, divided by $\hbar F$, is
\[
 \begin{aligned}
 \epsilon_J&=\mathcal J[\widetilde\eta]-\mathcal J[X_1+\chi_2]\\
 &=2\operatorname{Re}\langle X_1+\chi_2,mR_2\rangle+\langle R_2,mR_2\rangle,\\
 |\epsilon_J|&\le |2\operatorname{Re}\langle X_1,mR_2\rangle|\\
 &\quad+\sum_{m,\nu}\int r\,dr\,|m|(2|\chi_2||R_2|+|R_2|^2).
 \end{aligned}
\]
At the plus endpoint, $\epsilon_J=4.43\times10^{-6}$ and the absolute bound is $1.50\times10^{-4}$. Symplectic orthogonality removes the selected linear cross term; its measured magnitude below $3.7\times10^{-16}$ is retained in the bound.

\paragraph{Microscopic preparation and numerical checks.}
Away from linear resonance, pair frequencies differ. The Hermitian cavity coefficient uses the mean of their resolvents in the same angular and axial sector:
\[
 \begin{split}
 \overline{\mathcal R}_{ij|\ell n}=\tfrac12\big[&
 (\omega_{ij}J-\mathcal K_{m_{ij},k_{ij}})^{-1}\\
 &+(-\omega_{\ell n}J-\mathcal K_{m_{ij},k_{ij}})^{-1}\big].
 \end{split}
\]
Here $(\omega_0,\omega_1,\omega_2,\omega_3)=(E_P,-E_D,-E_D,-E_B)$ are bare cavity energies; Eq.~\eqref{sm-eq:legs} instead fixes the outgoing frequency to $-\Omega$. The inverses coincide at resonance and use no bound-pole projection. Static responses use the fixed-phase $\mathcal L_+$ solve; collective translation is excluded by the lattice. The baseline has 384 radial finite-volume cells, $h=R/384=0.03468161$, and $|m|,|\nu|\le4$. A dealiased $18\times18$ transform evaluates the local nonlinearity; implicit midpoint uses $\Delta t=0.01$, with no absorber. Baseline angular, atom-norm and relative excitation-energy drifts are below $2.3\times10^{-13}$, $1.5\times10^{-9}$ and $2.4\times10^{-7}$, respectively. Spatial and temporal resolution are varied in the controls below.

\paragraph{Resolution controls for the amplifying phase.}
The three $T=1000$ controls fix $R,L$, seeds and phases; the half-step run reuses the initial field. On 768 radial cells, we independently recompute the core, modes, 21 quadratic responses and fixed-number correction, retaining the residual detuning $-1.3564\times10^{-5}$ without radius retuning or interpolation. The Fourier case zero-extends the initial field to $|m|,|\nu|\le6$, uses a dealiased $25\times25$ transform, and reads all 13 signed Kelvin modes. Table~\ref{sm-tab:strong-controls} uses each actual initial field.
\begin{table}[t]
\caption{Changes $100[X(T)/X(0)-1]$ in percent, $T=1000$, for the plus preparation. $W_{\rm core},S_{\rm core}$ use $r<0.8\xi$, $d=3$. }
\label{sm-tab:strong-controls}
\begin{ruledtabular}\begin{tabular}{lrrrr}
Case & $K_{\rm b}$ & $E_{\rm b}$ & $W_{\rm core}$ & $S_{\rm core}$\\
Baseline & $17.5734$ & $-11.9269$ & $10.4822$ & $-16.9591$\\
$\Delta t/2$ & $17.5743$ & $-11.9275$ & $10.4964$ & $-16.7434$\\
$n_r=768$ & $17.5267$ & $-11.8902$ & $10.3668$ & $-16.7804$\\
$|m|,|\nu|\le6$ & $17.5723$ & $-11.9295$ & $10.4884$ & $-16.9525$\\
\end{tabular}\end{ruledtabular}
\end{table}
Across the controls and four core fits, maximum endpoint differences from baseline are $0.0467$, $0.0367$, $0.1163$ and $0.2158$ percentage points for $K_{\rm b},E_{\rm b},W_{\rm core},S_{\rm core}$. All refined runs retain the bound in Eq.~\eqref{sm-eq:band-bound} above $14.00\%$. Fast oscillations give larger full-trace deviations, up to $0.00815$ and $0.01012$ for normalized $W_{\rm core}$ and $S_{\rm core}$.

\paragraph{Units and boundaries.}
For $^{87}$Rb with scattering length $100a_0$ and $\mu/h=1\,$kHz, $\xi=0.241144\,\mu$m, $N_\xi=1.81315$ and $F=1431.609$. Then $R=3.21149\,\mu$m, $L=30.3030\,\mu$m, $T=159.155\,$ms, and $\xi|w_D^+(T)-w_D^-(T)|=45.67\,$nm. The axial cell is periodic. The selected-mode group velocities $(v_P,v_D,v_B)=(1.33198,0.73179,0.80541)$ give travel distances $(321,176,194)\,\mu$m over this interval. Buffers of $321\,\mu$m per side suggest a $670\,\mu$m uniform segment; this estimate uses the selected-mode group velocities. Preparation, end reflections and detection impose additional experimental requirements. 

\section{A conditional relation between energy and geometry}
\label{sm-sec:populations}

An initially sound-free, weak Kelvin packet has energy $E_0$ and axial momentum $P_0$. The remaining fundamental sector has $E_g=\sum_k E_K(k)n_k$ and $P_K=\sum_kkn_k$, in units of $\mu$ and $\hbar/\xi$; the cavity quantity $E_{\rm b}$ in Sec.~\ref{sm-sec:free} is instead energy per $\mu F$. Assume fixed background axial winding, no external momentum sink and additive weak-mode energy.

If all other carriers obey $e\ge c|k|$, their energy $E_0-E_g$ bounds their net momentum: $E_0-E_g\ge c|P_0-P_K|$. Thus
\begin{equation}
 P_K\ge P_{\min}=P_0-(E_0-E_g)/c.
 \label{sm-eq:momentum}
\end{equation}
For an axial period $L$, allowed wave numbers form the discrete set $k=2\pi j/L$, $j\in\mathbb Z$. Assume the allowed even Kelvin energies admit a convex interpolation $E_{\rm cvx}(k)$, nondecreasing for $k\ge0$. For $\mathcal N_K=\sum_kn_k$ and $P_{\min}>0$, Jensen's inequality gives
\[
 \begin{aligned}
 \frac{E_g}{\mathcal N_K}
 &=\sum_k\frac{n_k}{\mathcal N_K}E_{\rm cvx}(k)\\
 &\ge E_{\rm cvx}\!\left(\frac{P_K}{\mathcal N_K}\right)
 \ge E_{\rm cvx}\!\left(\frac{P_{\min}}{\mathcal N_K}\right).
 \end{aligned}
\]
The first inequality uses convexity; the second uses monotonicity and $P_K\ge P_{\min}>0$. For $E_K=\Lambda k^2$ with constant $\Lambda>0$, this becomes $\mathcal N_K\ge\Lambda P_{\min}^2/E_g$. At fixed positive $P_{\min}$, lowering $E_g$ raises the lower bound on $\mathcal N_K$. During evolution, $P_{\min}$ also varies with $E_g$. If $Q(k)\ge2$ on the occupied branch, Eq.~\eqref{sm-eq:geometry} additionally gives
\begin{equation}
 E_g\ge\mathcal N_K E_{\rm cvx}(P_{\min}/\mathcal N_K),
 \qquad W_K\ge2\mathcal N_K/F.
 \label{sm-eq:jensen}
\end{equation}
Here $\mathcal N_K=\sum_k n_k$ includes both axial directions. Convex straight segments join the allowed energies. The value $E_K(0)=0$ serves as a mathematical endpoint of the interpolation; collective translation is treated separately. The argument also applies to ensemble-mean occupations.

These bounds constrain the allowed states; they do not require the Kelvin number to grow monotonically.

\begingroup
\makeatletter\let\@FMN@list\@empty\makeatother
\let\smSavedLabel\label
\renewcommand{\label}[1]{\smSavedLabel{sm-bib-#1}}
\makeatletter
\def\bibliography@latex#1{\@input@{\jobname.bbl}}
\makeatother

\endgroup

\end{document}